\documentclass[twocolumn,showpacs,preprintnumbers,amsmath,amssymb]{revtex4}

\usepackage{graphicx}
\usepackage{dcolumn}
\usepackage{bm}

\begin{document}

\preprint{APS/123-QED}

\title{Low-temperature scanning tunneling microscopy of ring-like surface electronic structures around Co islands on
InAs(110) surfaces}

\author{D.~A.~Muzychenko, S.~V.~Savinov, N.~S.~Maslova,  V.~I.~Panov}

\affiliation{Faculty of Physics, Moscow State University,  119992
Moscow, Russia}

\author{K.~Schouteden, C.~Van~Haesendonck}
\affiliation{ Laboratory of Solid-State Physics and Magnetism,
BE-3001 Leuven, Belgium}

\date{\today}

\begin{abstract}
We report on the experimental observation by scanning tunneling
microscopy at low temperature of ring-like features that appear
around Co metal clusters deposited on a clean (110) oriented
surface of cleaved p-type InAs crystals. These features are
visible in spectroscopic images within a certain range of negative
tunneling bias voltages due to the presence of a negative
differential conductance in the current-voltage dependence. A
theoretical model is introduced, which takes into account
non-equilibrium effects in the small tunneling junction area. In
the framework of this model the appearance of the ring-like
features is explained in terms of interference effects between
electrons tunneling directly and indirectly (via a Co island)
between the tip and the InAs surface.
\end{abstract}

\maketitle

\section{Introduction}
After the pioneering work of Tsui \cite{Thui_70}, two-dimensional
(2D) electron systems have been intensively investigated, in
particular because of their huge application potential for the
manufacturing of field effect transistors. Since then, a lot of
success has been obtained with respect to improving the stability
and the carrier mobility of the 2D electron gas. On the other
hand, the 2D electron gas provides an excellent playing ground for
studying the physical properties of low-dimensional systems. Up to
now, the vast majority of experiments have been performed relying
on tunneling contacts with planar geometry, where data is
inevitably averaged and no local (on nanometer scale) information
can be obtained. The nanometer size tunneling contact, which is
created in scanning tunneling microscopy (STM) and scanning
tunneling spectroscopy (STS) experiments, is able to directly
provide such local information about the 2D electron gas.

Here, we present the results of local measurements of the
electronic properties of a 2D system, which consists of metal Co
islands deposited {\it in situ} on the (110) oriented surface of
InAs that is obtained by cleavage in ultra-high vacuum (UHV) of a
single crystal. It is well known that by deposition of only a
small amount of different adsorbates onto the InAs(110) surface,
the Fermi level can be shifted into the conduction band by up to
$600 \, {\rm meV}$ (see, e.g., \cite{Betti, Wiesendanger_Co,
Hasegava_2002} and references cited therein), thus causing strong
downward band bending and creating a 2D surface electron channel.

\section{Experiment and results}
\subsection{Experiment}
We performed combined STM and STS measurements with a commercially
available low-temperature STM \cite{Omicron}. The STM operates
under UHV conditions at a base pressure in the $10^{-12} \, {\rm
mbar}$ range and at temperatures down to $4.5 \, {\rm K}$. In
order to improve measurement stability, the whole UHV setup is
decoupled from the building by a specially designed vibration
isolation floor. Electrochemically preformed tungsten tips
\cite{Omicron}, which are flash annealed {\it in situ} at elevated
temperatures, are used. The tip quality is routinely checked by
acquiring atomic-resolution images for the ``herringbone"
reconstruction of the Au(111) surface.

Spectroscopic measurements were performed in the current imaging
tunneling spectroscopy (CITS) mode \cite{Wiesend_book} with a grid
size up to $200 \times 200$ points. Additionally, we relied on
harmonic detection with lock-in amplification for ``hardware"
measurements of the differential conductance. The excitation
frequency was in the range $300-7000 \, {\rm Hz}$, while the
excitation amplitude was in the range $5-100 \, {\rm mV}$.

The investigated InAs samples are doped with Mn at a doping level
of $5 \times 10^{17} \, {\rm cm}^{-3}$. At this doping level Mn in
InAs is expected to act as a shallow acceptor with ionization
energy around $28 \, {\rm meV}$. InAs slabs with size $5 \times 2
\times 2 \ \, {\rm mm}^3$ were cleaved {\it in-situ} at room
temperature in the UHV preparation chamber (base pressure is about
$5 \times 10^{-11}\, {\rm mbar}$) that is attached to the UHV STM
chamber. Co islands were deposited by means of electron-beam
evaporation at a rate of $0.007 \pm 0.001$ monolayer (ML) per
second. The evaporation rate was calibrated by determining with
x-ray reflectometry the thickness of thick Co films. As discussed
in more detail below, the evaporation rate inferred from STM
images of the Co islands on the InAs surface can deviate
considerably from the calibrated rate.

For the InAs(110) surface, filled and empty surface bands are
located deep inside the conduction and valence bands,
respectively, and are separated by a $1.7 \, {\rm eV}$ surface gap
\cite{Bulk_and_surface_1990, Bulk_and_surface_1998} for a clean
surface. Under clean conditions the bands are either flat, or a
weak accumulation layer is formed on the surface
\cite{Clean_accum}. However, surface states can be strongly
affected within 2 to 3 hours when the base pressure exceeds $(2-3)
\times 10^{-10} \, {\rm mbar}$ \cite{SS_sat_97}. In order to
prevent any undesirable effects of adsorbate induced surface
degradation, we always keep the pressure below $10^{-10}\, {\rm
mbar}$ in the preparation chamber of our UHV setup during the
different sample manipulations.

All the reported STM/STS experiments are performed at liquid
helium temperature ($\simeq 4.5 \, {\rm K}$). Everywhere in the
text the tunneling bias voltage $V_t$ refers to the sample
voltage, while the STM tip is virtually grounded. The experiments
consist of three stages. First, the clean surface of a cleaved
p-type Mn doped InAs(110) single crystal is characterized in
detail by STM/STS measurements. Second, Co metal atoms are
deposited on the InAs(110) substrate that is still cold (well
below room temperature) after removal from the STM chamber. The
amount of Co deposited on the surface is about $0.02 \, {\rm ML}$.
Deposition of a monolayer is according to the definition given in
\cite{Wiesendanger_Co}. Third, the Co-InAs system is investigated
by STM/STS measurements.

\subsection{2D surface of Co-InAs}
\begin{figure}
\includegraphics{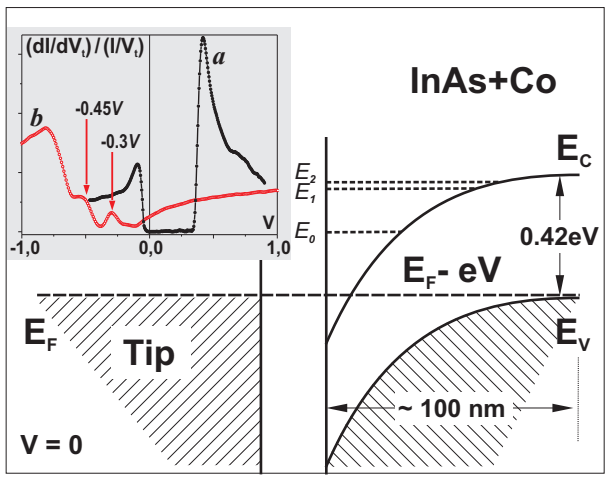}
\caption{\label{BB} Schematic diagram of the surface band bending
at the p-type InAs(110) surface after Co metal cluster deposition.
The band gap value, the width of the depletion layer and the
position of the 2D subbands are shown. Experimental normalized
tunneling conductance curves are shown in the inset both for the
original, clean p-type InAs(110) substrate and for the n-type Co
covered substrate (Co-InAs system). \newline Color online.}
\end{figure}

The quality of the clean cleaved InAs(110) surface is first
carefully checked. Stable STM images of the surface reveal (1x1)
relaxation with typical ``butterfly" type of patterns appearing at
the Mn doping atom sites \cite{AMS_Mn}, superimposed upon the InAs
atomic lattice (see Fig.~\ref{LockInImages}(a)). Mn atoms can be
visualized in the first few subsurface layers. Our STS
measurements clearly confirm the p-type character of the surface
conductivity (see inset in Fig.~\ref{BB}, curve a), as expected
for Mn doped InAs. After deposition of a very small amount of Co
atoms (only $0.02 \, {\rm ML}$) onto the surface, the general
outlook of the normalized ($dI/dV_{t})/(I/V_{t}$) tunneling
conductance curves drastically changes (see inset in
Fig.~\ref{BB}, curve b). The Fermi level $E_{F}$ is now positioned
about $0.12 \, {\rm eV}$ above the conduction band bottom. Two
distinct peaks are visible on the normalized tunneling conductance
curves around $-0.30 \, {\rm eV}$ and around $-0.45 \, {\rm eV}$.

In order to provide a consistent explanation for the observed
tunneling conductance curves, we need to take into account surface
band bending (see Fig.~\ref{BB}). Co clusters are acting as
``donors" when grown on a clean InAs(110) surface
\cite{Wiesendanger_Co}. The Co clusters are positively charged and
downward band bending occurs at the surface. We note that the
average value of the positive charge residing on a cluster is not
an integer multiple of a single electron charge \cite{Mas_Hab}.

Since our results indicate that $E_F - E_C \simeq 0.12 \, {\rm
eV}$, with $E_C$ the conduction band edge, and keeping in mind
that the band gap for InAs is $0.42\ {\rm eV}$ at $4.5 \ {\rm K}$
\cite{InAs_bandgap}, we then obtain a band bending at the surface
that is as large as $0.54 \, {\rm eV}$. This gives rise to a deep
quantum well with a few discrete eigenstates. The binding energy
of the lowest eigenstate (obtained by solving the
Poisson-Schr\"{o}dinger equation following \cite{Betti}) is $E_0
\simeq 0.24 \, {\rm eV}$. Consequently, the eigenstate is located
almost $0.30 \, {\rm eV}$ above $E_C$, or $0.18 \, {\rm eV}$ above
$E_F$. To make this state visible in the tunneling conductance
curve, it has to be shifted inside the tunneling window $[E_{F},
E_{F} - eV_{t}]$, and this can be achieved by making the tunneling
bias voltage sufficiently negative. When the eigenstate $E_0$
reaches $E_F$, a peak will appear in the normalized tunneling
conductance curve. We assume that this happens at a sample bias
voltage of about $-0.30 \, {\rm V}$ for our experiments. The next
peak appearing in the normalized tunneling conductance curve at
$V_t \simeq \ -0.45 \, {\rm V}$ can be assigned in a similar way
to the fact that the eigenstate $E_1$ falls below $E_F$. This is
similar to what was described for the GaAs(110) surface in
\cite{Feenstra_nGaAs}.

The top of valence band, $E_{V}$, is located $0.54 \, {\rm eV}$
below the Fermi level. Consequently, at bias voltage $V_t \, =
-0.45 \, {\rm eV}$ the width of the depletion layer is relatively
small, and a tunneling current is able to flow from the InAs
valence band to the tip. This can explain why the peak at $V_t
\simeq \ -0.45 {\rm V}$ is superimposed on a wide shoulder that
results from the bulk valence band states (see inset in
Fig.~\ref{BB}, curve b)

\subsection{Rings around Co clusters}
\begin{figure*}
\includegraphics{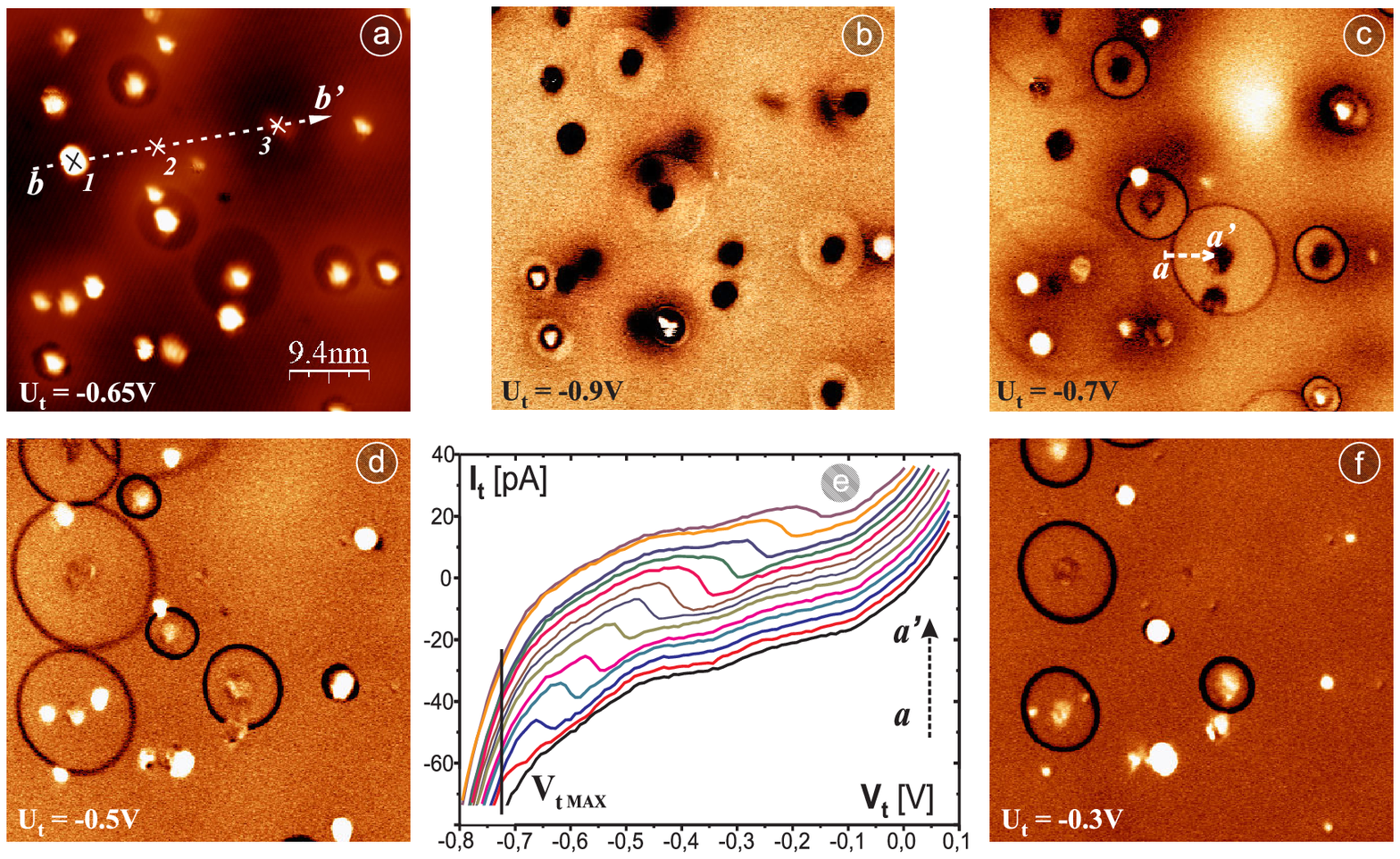}
\caption{\label{LockInImages}(a) Experimental constant current
topographical STM image of the Co-InAs(110) surface with a Co
coverage of about $0.02 \, {\rm ML}$ obtained at  $V_t = -0.65 \,
{\rm V}$ and $I = 400 \, {\rm pA}$. The image size is $47 \times
47 \, {\rm nm}^2$. (b), (c), (d) and (f) are maps of the measured
differential conductance ${dI/dV_t(x,y,V_t)}$ for the same surface
region as in (a) at different tunneling bias voltage. The
tunneling current setpoint is $I_t = 150 \, {\rm pA}$, while the
tunneling bias voltage is indicated at the bottom of the maps.
Ring-like features, whose appearance is strongly affected by the
applied negative bias voltage, can be observed. (e) Set of
$I(V_t)$ curves extracted from the CITS data (see text) along the
line $aa'$ (see (c)). The different curves are offset for clarity.
Regions with negative differential conductance are clearly
visible. The value of the tunneling bias voltage corresponding to
the maximum observable ring radius (see text) is marked as
$V_{tMAX}$. The ring in (c) that is below the line $aa'$ has
almost reached its maximum radius.}
\end{figure*}

According to the results of our STM measurements, Co atoms tend to
form small clusters on the InAs(110) surface despite the low
substrate temperature during deposition. A typical STM image of a
Co-InAs surface is presented in Fig.~\ref{LockInImages}(a). From
Fig.~\ref{LockInImages}(a) we infer that the average cluster size
is around $1.5 \, {\rm nm}$ for a tunneling bias voltage of $-0.65
\, {\rm V}$. It is important to explicitly refer to the value of
the tunneling bias voltage when analyzing topographical STM
images, because the apparent size of the Co clusters strongly
depends on the applied bias voltage. The majority of the clusters
are one monolayer in height. A rough estimate gives an average
number of Co atoms in a cluster (assuming a Co atom diameter of
about $0.25 \, {\rm nm}$) that is around 30 atoms. There are 21 Co
clusters on the STM image in Fig.~\ref{LockInImages}(a), which has
a size of $4.7 \times 4.7 \, {\rm nm}^2$. Consequently, the
coverage that is inferred from Fig.~\ref{LockInImages}(a) is
$0.068 \, {\rm ML}$, i.e., 3 to 4 times larger than the coverage
obtained from the calibrated deposition rate.

A very intriguing observation in our STM images of the Co islands
on the InAs(110) surface is the observation of ring-like features
that appear around the Co clusters. In Fig.~\ref{LockInImages}(a)
one observes more or less circular depressions with sharp edges
around some of the Co clusters. A surprising fact about these
depressions is that the depressions have a different diameter at
the same value of the bias voltage for clusters that look very
similar in the STM image. In previous STM experiments
\cite{Van_Kempen_96}, similar features were observed around metal
clusters of unknown origin.

In order to obtain a more detailed insight into the appearance of
the ring-like features, we performed spatially resolved
spectroscopic measurements above the same surface area where the
STM image was obtained (Fig.~\ref{LockInImages}(a)). In a first
set of experiments, we relied on imaging based on harmonic
detection with a lock-in amplifier. The results are shown in
Figs.~\ref{LockInImages}(b), (c), (d) and (f). Each image
corresponds to a map of the tunneling current response at the
frequency that is used for the modulation of the tunneling bias
voltage. In a first approximation, this response is proportional
to the value of the local density of states (LDOS) $\rho$
according to the relation
\begin {equation}
{dI/dV_t(x,y,V_t)} \propto \rho(x,y,E - E_F = e V_t) \: .
\nonumber
\end{equation}
Next, we performed high-resolution CITS measurements with grid
size of $130 \times 130$ points. Figure~\ref{LockInImages}(e)
shows a set of $I(V_t)$ curves acquired in the CITS mode along the
cross-section $aa'$ indicated in Fig.~\ref{LockInImages}(c).
Regions with negative slope, which corresponds to a negative
differential conductance (NDC), can be clearly observed on all but
the two lowest curves. The latter two curves were acquired near
point $a$, i.e., at points that are at the largest distance from
the cluster. The closer to the cluster the $I(V_t)$ dependence is
measured, the lower in absolute value is the tunneling voltage at
which the region with NDC occurs. This experimental finding
remains intact for STM tip to cluster distances down to a few
tenths of $1 \, {\rm nm}$. Comparing Fig.~\ref{LockInImages}(e)
with Figs.~\ref{LockInImages}(b), (c), (d) and (f), we come to the
conclusion that the distance $d$ between the STM tip and the
cluster determines the bias voltage $V_{NDC}(d)$ around which the
NDC occurs on the $I(V_t)$ curve. A dark ring with radius $d$,
which can be directly related to the presence of the NDC occurring
around $V_{NDC}(d)$, will appear on the ${dI/dV_t(x,y,V_t)}$ image
taken near the bias voltage $V_{NDC}(d)$.

From our analysis of the results shown in Fig.~\ref{LockInImages}
we conclude that (i) sharply defined dark rings appear in the
differential conductance (LDOS) images around some of the clusters
within a certain range of negative tunneling bias voltages, (ii)
the size of the dark rings is shrinking when the absolute value of
the tunneling bias voltage is decreased, and (iii) the LDOS has an
almost constant value above the Co-InAs surface except for the
clusters and the immediate vicinity of the dark rings. The only
image, which reveals pronounced spatial variations of the
normalized tunneling conductance, is image
Fig.~\ref{LockInImages}(c), taken at the threshold value of the
tunneling bias voltage $V_{t \, MAX}$ that approximately
corresponds to the bias voltage where the ring radius reaches a
maximum (see Fig.~\ref{LockInImages}(e)). We assume that the
presence and the different diameters of the dark rings around the
different clusters is caused by the different bonding of the
clusters to the InAs substrate. This assumption is supported by
the fact that different Co clusters look quite differently on STM
images taken at the same value of the tunneling bias voltage.
$I(V_t)$ curves also strongly differ when measured above different
clusters. Curves, which are measured above clusters that are
surrounded by a ring-like feature, reveal the presence of a narrow
plateau (width of about $0.1 \, {\rm V}$) around the Fermi level.
On the other hand, clusters, which are not surrounded by a
ring-like feature, reveal the presence of a wider plateau in the
$I(V_t)$ curve (width is $0.5-0.7 \, {\rm V}$). These findings
clearly require further investigation.

The maximum observable ring radius is determined by the fact that
the region with negative differential conductance (see
Fig.~\ref{LockInImages}(e)) becomes masked by the rapidly growing
tail of the $I(V_t)$ characteristics at more negative bias
voltages. For the Co clusters in the selected area this masking
occurs for voltages $V_t \lesssim V_{tMAX} \simeq -0.7 \, {\rm V}$
while for other areas the value of $V_{tMAX}$ will be different.
In this bias voltage range the behavior of the tunneling
conductance becomes much more complicated. As can be seen in
Fig.~\ref{LockInImages}(b), ring-like protrusions are observed in
the LDOS image rather than ring-like depressions. Some of the
rings become partly dark and partly bright. For these more
negative bias voltages, the exact relation between the local slope
of the $I(V_t)$ curve and the depression of the tunneling current
becomes important. More subtle details, such as the presence of
neighboring clusters, impurity atoms or surface defects, will then
have a pronounced influence on the imaging of the clusters.

\begin{figure}
\includegraphics{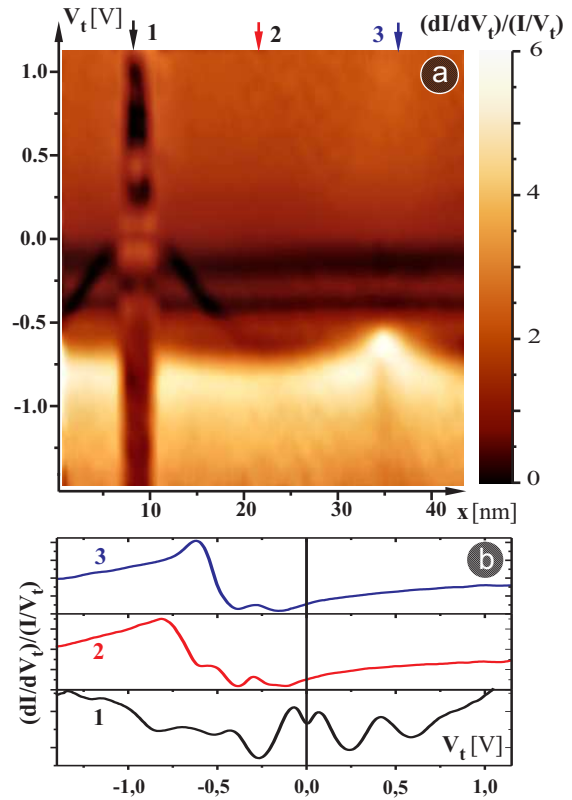}
\caption{\label{dI_slice} (a) 2D representation of the normalized
differential tunneling conductance, extracted from the CITS data
(see text) along the line $bb'$ indicated in
Fig.~\ref{LockInImages}(a). The horizontal axis corresponds to the
spatial coordinate, while the vertical axis corresponds to the
tunneling bias voltage. The color contrast in the image
corresponds to the value of the differential tunneling conductance
as given by the color scale on the right hand side. (b) Three
normalized differential conductance curves measured at the three
points (1, 2, and 3, see (a)) that are marked by the crosses along
the line $bb'$ in Fig.~\ref{LockInImages}(a). The curves are
measured above a cluster (1), above a defect free surface area (2)
and above a Mn impurity atom (3).}
\end{figure}

The large grid size that has been used for acquiring the CITS data
allows us to present the data in another interesting way.
Figure~\ref{dI_slice}(a) gives a 2D map of the variation of the
normalized tunneling conductance data along the cross-section
$bb'$ in Fig.~\ref{LockInImages}(a). The horizontal axis
corresponds to the spatial coordinate, while the vertical axis
corresponds to the tunneling bias voltage, and the color intensity
corresponds to the measured value of the normalized tunneling
conductance (LDOS). In Fig.~\ref{dI_slice}(b) we show
$(dI/dV_{t})/(I/V_{t})$ at the three points that are marked by the
crosses on the cross-section $bb'$ in Fig.~\ref{LockInImages}(a).
These three points (1, 2, and 3) are marked in
Fig.~\ref{dI_slice}(a) as well.

From Fig.~\ref{dI_slice}(a) it is clear that a Co cluster affects
the LDOS in the conduction and valence bands only very locally.
Except for a limited voltage range near $V_{t} = 0$, the influence
of a cluster on the LDOS is very rapidly decaying outside the
cluster. Inside the cluster, the conductivity in the valence band
is considerably suppressed. Some of the energy levels of the Co
cluster become visible in the voltage range corresponding to the
band gap and the conduction band. In Fig.~\ref{dI_slice}(a) there
are two tilted dark lines that approximately start from the peak
in the normalized tunneling conductance of the cluster and that
extend over a distance of about $8 \, {\rm nm}$ into the defect
free surface area. These two dark lines directly reflect the
presence of the dark rings around the clusters as well as the
dependence of the ring diameter on the tunneling bias voltage,
which is clearly non-linear.

The two peaks in the normalized tunneling conductance at $-0.3 \,
{\rm V}$ and at $-0.45 \, {\rm V}$ (see curve (a) in the inset to
Fig.~\ref{BB}) also appear in Fig.~\ref{dI_slice}(a) in the defect
free surface area. The peaks are visible as perfectly straight,
horizontal stripes. This implies that there occurs no noticeable
change in the ``rigid"  (both bands move together) band shifts
relative to the original bands of the p-type InAs. This is rather
surprising, because a charged cluster and a charged impurity atom,
separated by a distance of about $20 \, {\rm nm}$, are surrounding
the scanned area, and one might therefore expect changes to occur
in the measured band shift. The absence of changes in the rigid
band shift may be accounted for in the following way. First, both
the Co cluster and the Mn impurity atom can be treated as point
charges. It is known that a point charge potential does not
strongly affect the band bending, but rather gives rise to fine
structure in the LDOS~\cite{Mas_Hab}. Second, the variations of
the LDOS around the Co cluster are strongly confined in space,
while the Mn impurity atom alters the LDOS mostly near the top of
the valence band. Hence, the conduction band remains almost
undisturbed, with the conduction band bottom $E_C$ appearing as a
straight line in Fig.~\ref{dI_slice}(a). When the tunneling bias
voltage $V_t$, which is sweeping during the measurements of the
tunneling conductance, reaches the band gap region and goes down
to the top of the valence band, the energy levels $E_0, E_1$ of
the 2D quantum well become populated. Taking into account the very
high electron mobility along the surface \cite{InAs_2D_mobility},
we may assume that in this bias range perfect screening by the 2D
electron gas is taking place, hiding other details in the
electronic structure.

On the other hand, when the 2D subband $E_1$ is shifted below the
Fermi level, the tunneling bias dependent band bending noticeably
slows down. This can be accounted for by the overlapping of the
higher 2D subbands, which form a relatively wide continuum of
states, thus pinning the Fermi level in this continuum. In this
case the surface band structure ``stabilizes", i.e., an increase
of the absolute value of the negative bias voltage leads only to
slight changes in 2D subband energy.

Another important observation related to Fig.~\ref{dI_slice}(a) is
the influence of the Mn impurity atoms on (below) the Co-InAs(110)
surface on the tunneling spectra. In the vicinity of a Mn atom the
LDOS is affected deep inside the conduction and valence bands (not
visible in Fig.~\ref{dI_slice}(a)). Additionally, an intense peak
appears in Fig.~\ref{dI_slice}(a) near the valence band top and
most probably this peak reflects the energy position of the Mn
acceptor band. We note that the spatial extent of this peak is
relatively large, about $7 \, {\rm nm}$.

\section{Discussion}
In order to explain the observed effects consistently one needs to
take into account the presence of non-equilibrium and interference
effects in the small tunneling junction \cite{AMS_Charge}, which
can be described using the Keldysh diagram technique.

We need to explain two main peculiarities that are related to the
observation of the ring-like features. First, the presence of the
NDC region in the tunneling conductivity curves has to be
accounted for. Second, the limited spatial extent of the circular
features requires an explanation. Accordingly, this discussion
section has been divided into two subsections.

\subsection{Negative differential conductance}
We start from the following model Hamiltonian $\hat{H}$:

\begin{eqnarray}
            \label{H}
   \hat{H} &=& \hat{H}_{0} + \hat{H}_{tun} +\hat{H}_{cl} +
\hat{H}_{QW} \; ,  \nonumber
\end{eqnarray}

\noindent
where
\begin{equation}
     \hat{H}_{0} = \sum_{{\bf q},\sigma}
     (\varepsilon_q-\mu)c^+_{{\bf q} \sigma}c_{{\bf q} \sigma} + \sum_{{\bf p}, \sigma}
     (\varepsilon_p-\mu-eV)c^+_{{\bf p} \sigma}c_{{\bf p} \sigma}
     \nonumber
\end{equation}
describes the non-interacting electrons in the two electrodes: tip
states $({\bf p} \sigma)$, and bulk states $({\bf q} \sigma)$. The
part $\hat{H}_{cl}$ corresponds to the electron states in the
cluster:
\begin{eqnarray}
     \hat{H}_{cl} = \varepsilon_{a} \sum_{\sigma}
                 a^+_{\sigma}a_{\sigma} \; , \nonumber
\end{eqnarray}

\noindent where the operator $a_\sigma$ ($a^+_{\sigma}$) destroys
(creates) an electron in the cluster with spin $\sigma$. The part
$\hat{H}_{QW}$ describes the electron states in the quantum well
resulting from the Co cluster deposition and the STM tip induced
band bending:
\begin{eqnarray}
     \hat{H}_{QW} = \sum_{{\bf k}, m, \sigma}{\varepsilon_{{\bf k} m \sigma} c_{{\bf k} m \sigma} c^+_{{\bf k} m
     \sigma}} \; ,
\nonumber
\end{eqnarray}

\noindent where $c_{{\bf k} m \sigma}$ ($c^+_{{\bf k} m \sigma}$)
destroys (creates) an electron with spin $\sigma$ in the quantum
well, $m$ is the number labeling the 2D bands and $\bf k$ is the
2D wavevector. $\hat{H}_{tun}$ is responsible for tunneling
transitions from the quantum well and from the cluster into each
of the electrodes (tip or substrate):
\begin{eqnarray}
                 \label{H_tun}
  \hat{H}_{tun} =
    \sum_{ {\bf p},\sigma} T_{{\bf p} \sigma} c^+_{{\bf p} \sigma}a_{\sigma} +
   \sum_{ {\bf k},m,\sigma} T_{{\bf k} m} c^+_{{\bf k} m \sigma}a_{\sigma}
   + \nonumber\\
   \sum_{{\bf p} , {\bf k} , m , \sigma}W e^{i\varphi(r)} c^+_{{\bf k}m\sigma} c_{{\bf p}\sigma} + \sum_{{\bf q}, {\bf k}, m, \sigma} T_{{\bf q}}
   c^+_{{\bf k}m\sigma}
   c_{{\bf q}\sigma}
   + h.c.   \nonumber \; .
\end{eqnarray}

\noindent $T_{{\bf p} \sigma}$, $T_{{\bf k} m}$, $T_{{\bf q}}$ and
$W$ are the relevant tunneling matrix elements:
 $T_{{\bf k} m}$ for tunneling from the cluster level to the quantum well subbands,
 $T_{{\bf p} \sigma}$ for tunneling from the cluster level to the STM tip continuum,
 $T_{{\bf q}}$ for tunneling from the quantum well level to the bulk of the semiconductor,
 $W$ for tunneling from the quantum well level to the STM tip.

The differential tunneling conductance can then be written as
\cite{Hofstetter_2001, AMS_2002}

\begin{eqnarray}
\label{DI/DV}
 \frac{dI}{dV} \propto \Gamma \nu_k + \frac{2 \, \sqrt{\gamma_p
\gamma_k\Gamma \nu_k (1-\Gamma \nu_k)^2}}{(1+\Gamma\nu_k)^2} \,
\cos\varphi \: {\rm
Re} \, G^R(eV)  \nonumber\\
-\frac{2 \, \gamma_p \gamma_k ((1+\Gamma \nu_k)^2-4 \Gamma \nu_k
\cos^2\varphi)} {(\gamma_p + \gamma_k)(1+ \Gamma \nu_k)^3} \, {\rm
Im} \, G^R(eV) \nonumber \\ +\frac{2 \Gamma \nu_k (\gamma_k +
\gamma_p)}{(1 + \Gamma \nu_k)^3} \, {\rm Im} \, G^R(eV) \; , \; \;
\end{eqnarray}

\begin{figure}
\includegraphics{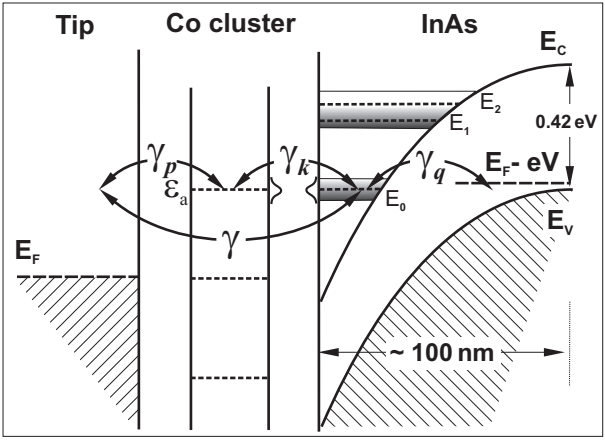}
\caption{\label{Model} Schematic representation of the tunneling
processes occurring in the system consisting of the STM tip, a Co
cluster and the InAs(110) substrate. $\varepsilon_a$ is the Co
cluster energy level that is participating in the tunneling. $E_0$
is the first 2D subband, which has a non-zero but narrow width, as
indicated by the gray stripe. A gray stripe is also used to mark
the presence of the higher 2D subbands, which form a continuum of
states due to their broadening. The system is shown at non-zero
bias voltage when $\varepsilon_a$ and $E_0$ are aligned and a dark
ring appears around the Co cluster in the maps of the differential
tunneling conductance.}
\end{figure}

\noindent were $\nu_k$, $\nu_p$ and $\nu_q$ are the density of
states in the quantum well, the STM tip and the bulk
semiconductor, respectively. The cluster density of states is
proportional to ${\rm Im} \ G^R(\omega)$.

The different tunneling rates in Eq.~(\ref{DI/DV}) are defined as
follows (see Fig.~\ref{Model}): $\gamma_k = T_{km}^2 \nu_{km}$ is
the tunneling transition rate from a cluster level to the quantum
well subbands, $\gamma_p = T_{p}^2 \nu_{p}$ is the tunneling
transition rate from a cluster level to the STM tip, $\gamma_q =
T_{q}^2 \nu_{q}$ is the tunneling transition rate from the quantum
well to the bulk of the semiconductor, and $\gamma = W^2 \nu_{p}$
is the tunneling transition rate from the quantum well to the STM
tip. Finally, $\Gamma$ is defined as $\Gamma = \gamma
\gamma_q/(\gamma + \gamma_q)$. We omit the subband index $m$,
because only the lowest subband plays the most important role in
our model.

As stated above, a Co cluster acts as a donor, and consequently it
is positively charged. In case of non-equilibrium, the energy of a
cluster level $\varepsilon_a$, which is participating in the
tunneling (see Fig.~\ref{Model}), depends on the electron filling
number $<n_a>$ of this level:
\begin{equation}
\varepsilon_a = \varepsilon_0 + U (n_0 \, - <n_a>) \; , \nonumber
\end{equation}
\noindent where $\varepsilon_0$ is the unperturbed position of the
cluster energy level, $n_0$ is the equilibrium cluster level
filling number, and $U$ is the Coulomb interaction energy of the
localized charges \cite{AMS_Charge}.

For STM junctions we can assume that $\Gamma \nu_k << 1$, because
$\Gamma$ is of order of tunneling rates and $\nu_k \sim 1/D$,
where $D$ is the width of conduction (valence) band in InAs.
Equation~(\ref{DI/DV}) then reduces to

\begin{eqnarray}
\label{DeltaDI/DV} \Delta (\frac{dI}{dV}) \propto 2 \mid {\rm Im}
\ G^R(\omega) \mid \frac{\gamma_p \gamma_k - \Gamma \nu_k
(\gamma_p + \gamma_k)^2}{\gamma_p +
 \gamma_k} + \nonumber \\
 2 {\rm Re} \ G^R(\omega) \sqrt{\gamma_p \gamma_k \Gamma \nu_k} \  \cos
 \varphi \; ,
\end{eqnarray}

\noindent where $\Delta$ refers the the extra contribution to the
tunneling conductance resulting from the non-equilibrium effects.

\noindent Here $\varphi$ is the pase angle, ascuired due to
difference of wavefunctions of direct (STM tip-2D subband) and
indirect resonant tunneling. The complicated coefficient in front
of ${\rm Im} \ G^R(\omega)$ and the presence of ${\rm Re} \
G^R(\omega)$ in Eq.~(\ref{DeltaDI/DV}) also results from
interference between direct tuneling from the STM tip to a quantum
well subband, and indirect resonant tunneling via the cluster
level. Consequently, the involved tunneling rates determine the
sign and the magnitude of the additional current resulting from
the non-equilibrium and interference processes.

For each tip position the value of $\gamma_p$ is fixed, while
$\gamma_k$ can be varied by changing the applied bias voltage.
$\gamma_k$ depends on the relative position of the cluster level
$\varepsilon_a$ and of the first 2D subband $E_0$. Both energies
are bias dependent: $\varepsilon_a$ trough $<n_a>$ and $E_0$
through the band bending. It is important to note that the 2D
subband $E_0$ is relatively narrow. Electrons can be injected into
this subband either from the STM tip ($\gamma$) or from the Co
cluster ($\gamma_k$). In both cases the wavevector component
parallel to the surface, $k_{||}$, is small. In close analogy with
angle resolved photoemission spectroscopy experiments
\cite{Aristov_99}, only a narrow range of energies is available
for tunneling into the 2D subband $E_0$ (see Fig.~\ref{Model}). In
general, the $k_{||}$ component needs to be taken into account in
STM experiments \cite{Ebert_momantum}.

We assume that the unperturbed Co cluster energy level
$\varepsilon_0$ (at zero bias voltage) is positioned close to
$E_F$ below the first 2D subband $E_0$, i.e, the two energy levels
are not aligned. This assumption is strongly supported by the
experimental STS data (see Fig.~\ref{dI_slice} for details).

We now proceed with a more detailed analysis of the bias voltage
dependence of $\Delta(dI/dV)$. The first term in
Eq.~(\ref{DeltaDI/DV}), which varies proportional to the cluster
density of states, can change its sign. If $\gamma_p \gamma_k <<
\Gamma \nu_k (\gamma_p + \gamma_k)^2$ (let us assume $\gamma_p <<
\gamma_k$) the sign of the coefficient that appears in front of
${\rm Im} \ G^R(\omega)$ is negative. For the assumed conditions,
the coefficient that appears in front of ${\rm Re} \ G^R(\omega)$
is much smaller than the coefficient next to ${\rm Im} \
G^R(\omega)$. The tunneling conductance curve is an inverted
Lorentzian-shaped curve that is slightly asymmetric because of the
${\rm Re} \ G^R(\omega)$ contribution. The position of the dip in
the tunneling conductance curve can be connected to the
$\varepsilon_a(<n_a>)$ dependence. The closer to the cluster the
STM tip is, the higher is the value of $\gamma_p$ and the lower is
the value of the non-equilibrium filling $<n_a>$. The cluster
energy level $\varepsilon_a$ is pushed higher, and the dip in the
tunneling conductance curve can be observed at smaller absolute
values of the applied bias voltage (see Fig.\ref{LockInImages}).
The dark rings that appear in the STS images shrink when the
absolute value of the negative tunneling bias decreases within a
cluster dependent range of negative voltages.

\subsection{Spatial extent of ring-like features}

The most probable physical mechanism causing the formation of
narrow (of the order of the interatomic distance) ring-like
features in real space is some kind of resonant tunneling. When
the STM tip changes its spatial position, localized energy levels
that participate in the tunneling processes are coming out of
resonance or even are falling outside the tunneling window
$[E_{F}, E_{F} - eV_{t}]$. We find that the simple cluster
charging model used in \cite{Van_Kempen_96,Rings_GaAs} is not
adequate for describing our Co-InAs system. This is mainly due to
the fact that on-site Coulomb repulsion for the cluster is the
strongest bias dependent effect. This leads to variations of the
cluster level filling number, and consequently to changes in the
cluster level energy. Starting from this finding we are able to
analyze in the framework of our model how spatially narrow
circular features can be formed around the Co clusters.

For each value of the tunneling bias voltage $V_t$ the value of
$\gamma_p$ can be altered by changing the STM tip to cluster
distance. Consequently, $<n_a>$, which depends on the ratio
$\gamma_k / \gamma_p$ also changes, giving rise to a shift of
$\varepsilon_a(<n_a>)$ and to variations of $\gamma_k$. When
moving the STM tip farther away from the cluster, $\gamma_p$ is
decreasing, leading to an increase of $<n_a>$ and to a lowering of
$\varepsilon_a$. At a certain distance $d_{max}$, the cluster
energy level $\varepsilon_a(<n_a>)$ starts to becoming out of
resonance with the narrow 2D subband $E_0$. Since there is now an
alignment with the tail of the broadened $E_0$ level, the value of
the tunneling current drops. At the same time, the value of
$\gamma_k$ decreases, resulting in a decrease of the
non-equilibrium occupation $<n_a>$, and consequently
$\varepsilon_a$ is pushed higher. The cluster level
$\varepsilon_a$ then jumps to its equilibrium position
$\varepsilon_0$. We assume that at the STM tip to cluster distance
$d_{max}$ the tunneling rate $\gamma_p$, which exponentially
decreases with increasing distance, becomes negligible ($\gamma_p
\approx 0$). The STM tunneling system does no longer feel the STM
tip - cluster interaction, and the value of the tunneling current
returns to its ``normal" value. In other words, when at a certain
tunneling bias voltage in the range where a dark ring appears in
the LDOS around the Co cluster, the STM tip to cluster distance
$d$ reaches the value $d_{max}$, some kind of positive feedback,
caused by mutually coupled variations of $\gamma_p, \gamma_k$ and
$<n_a>$, switches on. This positive feedback abruptly shifts the
cluster energy level out of resonance with the 2D subband. This
can account for the very small spatial width of the rings that
appear in the LDOS as well as for the non-linear dependence of the
ring diameter on the STM tip to cluster distance (see
Fig.~\ref{dI_slice}).

We note that for specific tunneling conditions the first term in
Eq.~(\ref{DeltaDI/DV}) can become positive, causing the
observation of bright rings or protrusions in the images of the
differential conductance. A dark ring can turn into a bright one
at high values of the tunneling bias voltage, as can be seen in
Fig.~\ref{LockInImages}(b).

Our theoretical model does not include any assumptions concerning
the physical origin of the localized states through which resonant
tunneling can occur. Therefore, this model should also be
applicable to the case of an individual impurity atom near the
surface. Ring-like features around individual impurities were
indeed observed in our experiments (to be published) as well as on
the GaAs(110) surface \cite{Rings_GaAs}.

\section{Conclusion}
In conclusion, relying on combined STM and STS measurements at low
temperature, we have been able to identify the presence of
ring-like features around Co metal clusters on a p-type InAs(110)
surface that is prepared by {\it in situ} cleavage of a single
crystal in UHV. The ring-like features become clearly visible in
the maps of the differential normalized conductance for a certain
range of negative tunneling bias voltages, due to the presence of
a region with negative differential conductance in the $I(V_t)$
dependence. The diameter of the rings is decreasing when
decreasing the absolute value of the tunneling bias voltage. The
possibility to observe the ring-like features around a Co cluster
depends on the bonding of the cluster to the InAs(110) surface.

A theoretical model was developed, which takes into account the
non-equilibrium effects that occur in the small STM tunneling
junction. The model we propose allows us to explain the main
experimental findings. In the framework of the model the
appearance of the ring-like features structures is accounted for
by the interference of direct (between the STM tip and the QW) and
indirect tunneling via a Co cluster.

\begin{acknowledgments}

This research in Moscow has been supported by the RFBR grants
05-02-19806-MF-a and 06-02-17076-a. The research in Leuven has
been supported by the Fund for Scientific Research - Flanders
(Belgium) as well as the Research Fund of the K.U.Leuven. We
thanks A.~A.~Ezhov for the technical support. The authors also
thank the WSxM team (www.nanotec.es).

\end{acknowledgments}

\bibliography{NT_DM08122007}

\end{document}